\documentclass{article}
\pdfoutput=1

\usepackage[english]{babel}

\usepackage[letterpaper,top=2cm,bottom=2cm,left=3cm,right=3cm,marginparwidth=1.75cm]{geometry}

\usepackage{amsmath}
\usepackage{graphicx}
\usepackage[colorlinks=true, allcolors=blue]{hyperref}

\title{\textbf{The Kinematic Distance to NGC 6309}}
\author{\textbf{Scott C. Scharlach,$^1$ Colton G. Morgan$^1$}}

\begin{document}

\maketitle

\begin{center}
{\textit{$^1$Pisgah Astronomical Research Institute \\
1 PARI Drive \\
Rosman, NC 28772, USA}}
\end{center}

\begin{abstract}
 We report an updated value for the distance to the planetary nebula NGC 6309 (the Box Nebula). The distance is found through two Kinematic Distance Methods (KDMs): the system of two equations reported in Zhu et al. 2013 and the Monte Carlo method reported by Wenger et al. 2018. We find the kinematic distance to NGC 6309 to be 4.1 kpc with an upper uncertainty of +0.29 kpc and a lower uncertainty of -0.38 kpc. We also calculate the distance to Cassiopeia A with the two KDMs and compare to the value reported by Reed et al 1995. The Zhu et al. method and Wenger et al. method yield a value within thirty percent and twenty percent of the Reed et al. method, respectively. The value reported by Reed et al 1995 was contained within the error bounds produced by the Wenger et al. method. The distance measurement to Cassiopeia A suggests that both KDMs, while imperfect, are moderately accurate methods for determining the distance to NGC objects in the plane of the Milky Way.

\end{abstract}

\begin{center}
\end{center}

\section{Introduction}

Planetary Nebulae (PNe) are highly luminous clouds of gas and dust around post-main-sequence stars. Stars which range in mass from 0.8 to 8 solar masses emit PNe prior to the star becoming a red giant. Because PNe are created at a transition period in star's life cycle, PNe provides an important source of information on the evolution of stars in the universe.

In order to determine several features of PNe, such as radius, age, mass, and luminosity, astronomers must first accurately measure the object's distance. However, the distances to PNe have historically been difficult to calculate. Methods such as those proposed in Shklovsky 1956 or Daub 1982 provide order-of-magnitude estimates, but they are imprecise.

Two recent astronomical discoveries have allowed for a different distance calculation, the Kinematic Distance Method, to be far more feasible than it was only a few decades ago. Firstly, Reid et al 2014 published an updated function between the galactocentric radius of objects in the Milky Way and the circular velocity of the object around the center of the Milky Way. Secondly, Abuter et al 2019 discovered a highly accurate distance to the center of the galaxy with an uncertainty of 0.3 $\%$. These two values allow astronomers to calculate the distance to PNe with greater accuracy than in the past.

This paper aims to discover the distance to one object, NGC 6309 (nicknamed the Box Nebula), in order to demonstrate the efficacy of the Kinematic Distance Method. The distances to PNe allows astronomers to calculate other important intrinsic features of PNe, such as their radius,  age, absolute magnitude, and luminosity. In other words, an accurate understanding of the distance to an object is essential for discovering a multitude of the object's intrinsic features.

\section{Methods}

\subsection{Kinematic Distance}
\label{Kinematic_Distances}

Objects within the Milky Way galaxy rotate in approximately circular orbits around the center of the galaxy. For objects with a galactocentric radius\footnote{The galactocentric radius of an object is the distance between the object and the center of the Milky Way galaxy.} of 6 kpc or greater, the function between galactocentric radius and circular velocity around the Milky Way is approximately flat. In other words, objects in the plane of the Milky Way with a galactocentric radius of 6 kpc or greater have roughly equal circular velocities.

When astronomers observe an object in the plane of the Milky Way, there is a component of the object's circular velocity which points directly away from the observer. The rate at which an object moves towards or away from an observer in our solar system is called the radial velocity of the object.

Using trigonometry, Brand and Blitz 1993 presented an equation which relates the object's circular velocity to its observed radial velocity. (The equation assumes the object has no proper motion, i.e. the object perfectly adheres to the galactic rotation curve function. The phenomenon of proper motion is discussed in Section 2.3.) The equation is:

\begin{equation} 
\label{eq:BandBredshift}
    V_r = (V_c\frac{R_0}{R} - V_0)sin(l)cos(b) ,
\end{equation}

where $V_r$ is heliocentric radial velocity of the object, $V_{c}$ is the circular velocity of the object around the center of the Milky Way, $R_0$ is the galactocentric radius of the Local Standard of Rest (LSR)\footnote{Additional information on the LSR is found in the Appendix.}, $R$ is the galactocentric radius of the object, $V_0$ is galactocentric velocity of the LSR, and $(l,b)$ are the galactic coordinates of the object.

Zhu et al 2013 approximated the circular velocity of the object and the circular velocity of the LSR as approximately equal. The equation then simplifies to:

\begin{equation} 
\label{eq:redshift}
    V_r = V_0(\frac{R_0}{R} - 1)sin(l)cos(b) .
\end{equation}

An object, the LSR, and the center of the galaxy form the points of a triangle. Thus, their relative distances are related via the Law of Cosines:

\begin{equation}
\label{eq:law of cosines}
   R^2 = R_0^2 +(dcos(b))^2 - 2R_0dcos(b)cos(l) ,
\end{equation}

where $d$ is the heliocentric distance to the object. The equation utilizes the term $dcos(b)$ to account for the angle between the plane of the Milky Way and the object, as the galactic object is not necessarily in the exact plane of the Milky Way.

Wenger et al 2018 reports that the circular velocity of the LSR ($V_0$) is defined to be 220 $km* s^{-1}$. Abuter et al 2019 reports the galactocentric radius of the LSR ($R_0$) to be 8.178 kpc, with an uncertainty of 0.3$\%$.

$V_r$ and $(l,b)$ are observable quantities which are specific to the observed object. In particular, $V_r$ can be determined from the object's redshift, which in turn can be determined from spectral data. The center of the galaxy has a known location in the celestial sphere; Wenger et al. 2018 reports the galactic coordinates of the center of the Milky Way to be $(0^{\circ}, 0^{\circ})$ by definition. The galactic coordinates $(l,b)$ can be determined simply by measuring the angle between the center of the galaxy and the object relative to the observer.

Because $V_0$, $R_0$, $V_r$, and $(l,b)$ are either known or observable values, equations \ref{eq:redshift} and \ref{eq:law of cosines} thus form a system of two equations with two unknowns, $R$ and $d$. This system can be solved to determine the heliocentric distance to objects such as planetary nebulae.

\subsection{The Kinematic Distance Ambiguity and its Resolution}

\begin{figure}[ht]
\centering
\includegraphics[width=0.3\textwidth]{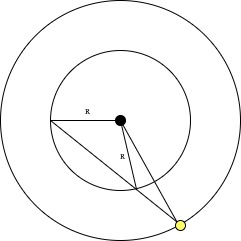}
\caption{\label{fig:The Kinematic Distance Ambiguity (KDA)} The small black disk indicates the center of the galaxy. The inner ring indicates the galactic orbital paths of a hypothetical object, while the outer ring represents the galactic orbital path of the solar system. The yellow disk indicates the solar system. The system of two equations described in the text provides not one but two possible heliocentric distances at the same galactocentric distance $R$ when $R < R_0$.}
\end{figure}

For objects with a galactocentric radius which is smaller than the galactocentric radius of the LSR, the system of two equations described by \ref{eq:redshift} and \ref{eq:law of cosines} provide two possible values for the heliocentric distance $d$, a near distance and a far distance. Figure \ref{fig:The Kinematic Distance Ambiguity (KDA)} demonstrates two possible locations of a hypothetical object which both satisfy the system of two equations. This uncertainty is called the Kinematic Distance Ambiguity (KDA). Methods which attempt to resolve the KDA are called the Kinematic Distance Ambiguity Resolution (KDAR).

This paper resolves the KDA with the distance estimation method originally reported by Daub 1982 and modified by Cahn et al 1991. Firstly, Daub 1982 presents an equation for the optical thickness parameter $T$, which is given by

\begin{equation}
\label{eq:thicc}
   T = log(\frac{4\theta^2}{F(5)}) ,
\end{equation}

Where $\theta$ is the angular radius in arcseconds and $F(5)$ is the 5 GHz radio flux in Janskys. Both $\theta$ $F(5)$ are observable quantities.

Secondly, Cahn 1991 defines an abstract value $\mu$:

\begin{equation}
\label{eq:mass}
   \mu = \sqrt{2.266*10^{-21}D^5\theta^3F(5)} ,
\end{equation}

where $D$ is the distance to the object.

\begin{figure}[ht]
\centering
\includegraphics[width=0.5\textwidth]{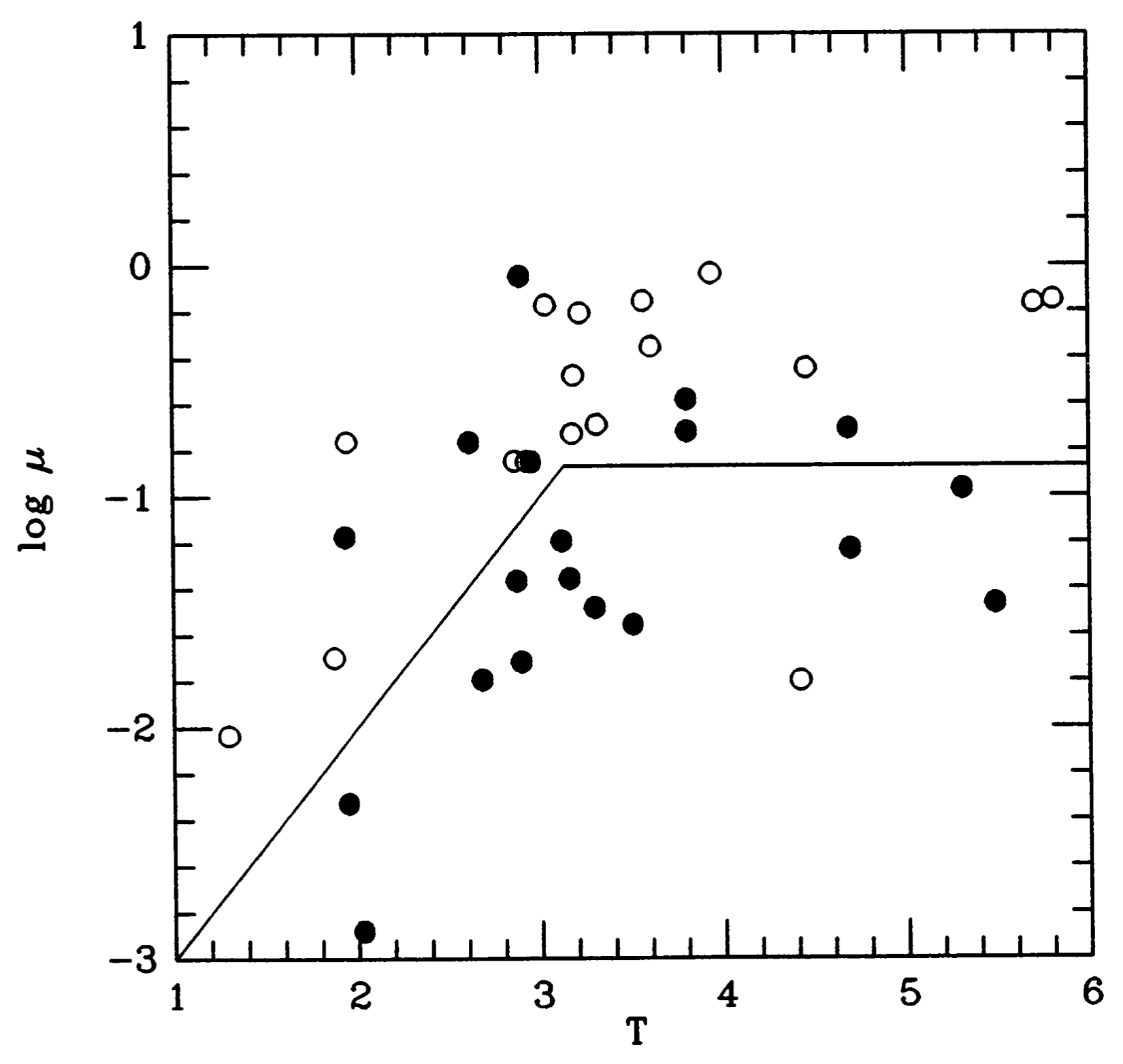}
\caption{\label{fig:Cahn} The optical thickness roughly correlates with the abstract quantity $\mu$, which is related to the object's heliocentric distance. Reprinted from Cahn et al 1991.}
\end{figure}

Thirdly, Cahn 1991 measured the distances to several PNe with the statistical parallax method, which allowed for the calculation of $\mu$ for those PNe. The publication presents a a rough, order-of-magnitude correlation between $log(\mu)$ and optical thickness $T$. Figure \ref{fig:Cahn} shows this approximate correlation. Cahn 1991 presumed that the correlation remains true for PNe which are too distant for accurate statistical parallax measurements.

Fourthly, by observing the 5 GHz radio flux of an object and its angular radius, astronomers can use equation \ref{eq:thicc}, equation \ref{eq:mass}, and Figure  \ref{fig:Cahn} to roughly estimate the distance to a planetary nebula, even when statistical parallax data is unavailable. However, the correlation shown in \ref{fig:Cahn} is imprecise, and therefore the distance estimate is imprecise as well.

Generally, only one of the two distance values from the system of two equations described above falls within the uncertainty range of the  method described in this section. The distance value which satisfies both methods is more likely to accurately reflect the distance to the object than the distance estimate which satisfies only one of the two methods.

\subsection{Proper Motion and a Monte Carlo Method}

The Kinematic Distance Method relies on an accurate understanding of the rate at which the object rotates around the center of the Milky Way. Astronomers determine this from observing the object's radial velocity. However, the object may also have proper motion, i.e. motion which does not perfectly adhere to the predicted rotation rate. Proper motion can affect the object's observed radial velocity to a non-negiligible degree.

Proper motion presents an important problem for the KDM because equations and  \ref{eq:BandBredshift} and \ref{eq:redshift} assume that the observed radial velocity is due only to galactic rotation, not proper motion. However, Brand and Blitz 1993 point out that there is no $\emph{a priori}$ method for determining what percentage of the observed radial velocity is due to galactic rotation and what percentage is due to proper motion.

The problem of proper motion can be partially accounted for with methods developed by Brandon and Blitz 1993 and Wenger et al 2018, both of which are described in this section.

Brandon and Blitz 1993 developed a method to calculate the proper motions of PNe with previously determined $d$ and $R$. (In this context, $d$ and $R$ were found by distance estimation methods which are independent of the Kinematic Distance Method.) Firstly, they inserted the galactocentric radius $R$ into an empirically derived equation for the rotation rate of the Milky Way at a given $R$: 

\begin{equation}
\label{eq:circular velocity}
   \frac{V_c}{V_0} = 1.00767(\frac{R}{R_0})^{0.0394}+0.00712 .
\end{equation}

Secondly, Brandon and Blitz substituted the calculated value of $V_c$ into equation \ref{eq:BandBredshift} in order to calculate the expected radial velocity due only to the circular velocity. Thirdly, they subtracted the actual observed radial velocity from the predicted radial velocity to find a value which the Brandon and Blitz 1993 referred to as the ``velocity residual.'' The velocity residual measures the degree to which the object's motion deviates from the galactic rotation curve; it is tantamount to the object's proper motion.

\begin{figure}[ht]
    \centering
    \includegraphics[width=0.5\textwidth]{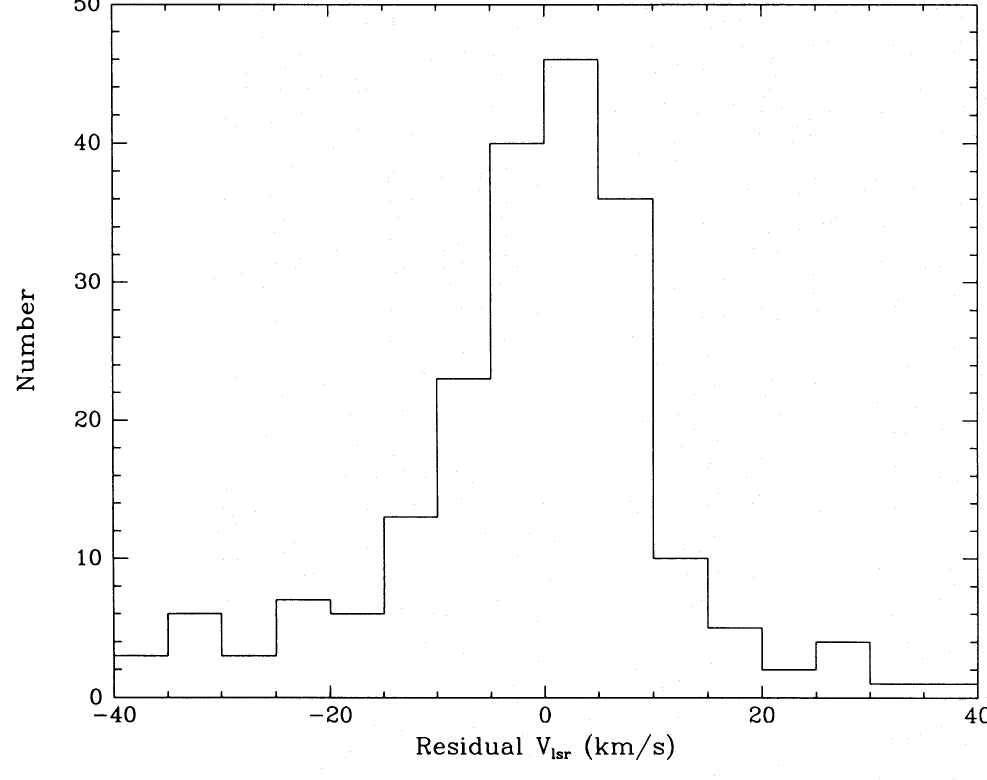}
    \caption{A histogram displaying the velocity residuals of planetary nebulae with $R$ and $d$ estimates. Some planetary nebulae have a velocity residual whose magnitude is large enough to create a systematic error in the distance estimation method described in Section \ref{Kinematic_Distances}. Reprinted from Brandon and Blitz 1993.}
    \label{fig:Histogram.png}
\end{figure}

Brandon Blitz presented a histogram which displays the velocity residuals of many planetary nebulae, which is reprinted in figure \ref{fig:Histogram.png}. Most PNe have velocity residuals which are 6.4 $km*s^{-1}$ or lower, but some have velocity residuals with an absolute magnitude as great as 40 $km*s^{-1}$.

Wenger et al 2018 presents open access software which accounts for the proper motion of PNe in a statistically robust manner. The software utilizes a Monte Carlo method in which each simulation randomly selects a proper motion for the object. The value of the proper motion is statistically weighted in a manner that matches figure \ref{fig:Histogram.png}. By conducting the simulation 10,000 times, the software produces a Gaussian curve of probable distances to an object with an observed $V_r$ and $(l,b)$.

The software also takes into account uncertainties in other variables which are beyond the scope of this paper. For example, the software does not approximate the galactic rotation curve as flat. Instead, it utilizes an updated, empirically-derived galactic rotation function from Reid et al 2014:

\begin{equation}
\label{eq:Reid}
   V_c(R)=241[\frac{1.97*\beta*x^{1.22}}{(x^2+0.78^2)^{1.43}}+(1-\beta)x^2\frac{1+1.46^2}{x^2+1.46^2}]^{1/2},
\end{equation}

where $x=R/(0.90R_0)$ and $\beta=0.72+0.44log_{10}[(1.46/1.5)^{5}]$.

By utilizing a Monte Carlo method with randomly selected proper motions and attending to precise details such as updated rotation curves, the software from Wenger 2018 allows astronomers to calculate the most likely distance to planetary nebulae even when the exact proper motion of the object is unknown.

\section{Results}

\subsection{The Distance to NGC 6309}

SIMBAD reports that the galactic coordinates $(l,b)$ of NGC 6309 are $(l,b)=(9.65484^{\circ}, +14.81371^{\circ})$.\footnote{SIMBAD was accessed via webpage through the following hyperlink: http://simbad.cds.unistra.fr/simbad/sim-id?Ident=NGC+6309\&NbIdent=1\&Radius=2\&Radius.unit=arcmin\&submit=submit+id}

Vázquez et al 2014 and Rubio et al 2014 both report the radial velocity of NGC 6309 to be $V_r = -32 \pm km*s^{-1}$. When this negative value for $V_r$ is 
inserted into the system of two equations, one distance estimate is negative and the other estimate is greater than the diameter of the Milky Way. The peculiar nature of these values suggest that the values are not physically meaningful, and instead the coordinate system ought to be redefined so that $V_r$ is positive. For this reason, we use $V_r = 32 \pm km *s^{-1}$ as our value for heliocentric radial velocity of NGC 6309.

The system of two equations yielded two distance values for NGC 6309: a near distance of 4.1 kpc and a far distance 12.6 kpc. Cahn 1991 estimates the distance to NGC 6309 to be 2.532 kpc, which is much closer to the near distance than the far distance. Therefore, the authors of this paper conclude that the near distance estimate, 4.1 kpc, is the more accurate distance estimate.

\begin{figure}[ht]
    \centering
    \includegraphics[width=1.0\textwidth]{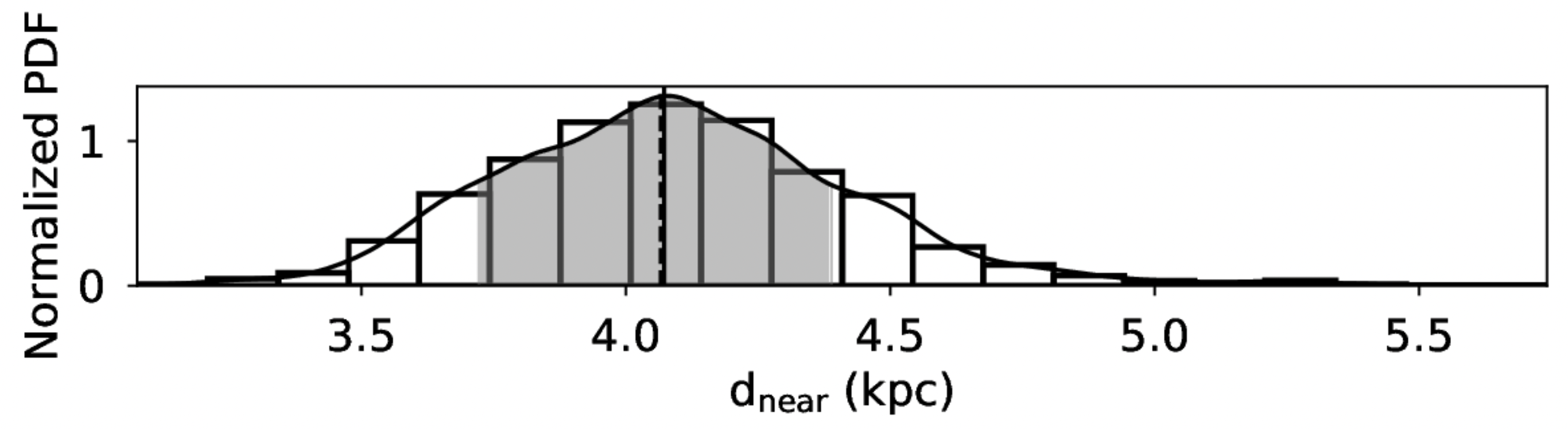}
    \caption{The software created by Wenger 2018 finds the distance to NGC 6309 to be 4.07 +0.32/-0.35 kpc, with the uncertainty indicating the 68.3$\%$ confidence interval.}
    \label{fig:PDF}
\end{figure}

The Wenger 2018 software produced a Probability Distribution Function (PDF) of the most likely range of distances to NGC 6309. The most likely distance to NGC 6309 is 4.07 kpc, and the 68.3$\%$ confidence interval ranges from a distance of 3.72 kpc to 4.39 kpc. The PDF takes the form of a Gaussian curve and is shown in figure \ref{fig:PDF}.

The Zhu 2013 method and Wenger 2018 method yielded distance estimates which differ only by 0.03 kpc; they are in excellent agreement with each other. This agreement in values suggests that the Kinematic Distance Method is replicable, which in the opinion of the authors adds to the method's reliability.

We use the value 4.1 kpc (with two significant figures) as our primary distance estimate to NGC 6309, and we use the 68.3$\%$ confidence interval from the PDF 
as our interval of uncertainty in our distance estimate. With these values, we conclude that the distance to NGC 6309 is 4.1 +0.29/-0.38 kpc.

\subsection{Efficacy Check: Cassiopeia A}

In order to check the reliability of the Kinematic Distance Method, we applied the KDM to an object with a known distance: the supernova remnant Cassiopeia A. Reed et al 1995 reported that the distance to Cassiopeia A is 3.4 kpc. Cassiopeia A was chosen for this efficacy check because it is in the plane of the Milky Way, it is has a documented radial velocity, and the distance estimate used by Reed 1995 was conducted independently of the Kinematic Distance Method.

Reed 1995 originally calculated the distance to Cassiopeia A with a five-step method. Firstly, the authors utilized the object's redshift to calculate its expansion velocity in $km*year^{-1}$. Secondly, the authors observed the object's change in angular radius over time, in $as *year^{-1}$ (arcseconds per year). Thirdly, the authors observed the object's angular radius in $as$. Fourthly, the authors used the three previous values to calculate the object's radius in $km$. Finally, the authors calculated the object's distance using the trigonometric equation $sin(\theta)=r/d$, where $\theta$ is the angular radius, $r$ is the radius, and $d$ is the heliocentric distance to the object.

With regard to the Kinematic Distance Method, Asgekar et al 2013 reports the radial velocity of Cassiopeia A to be approximately -48.0 $\pm$ 1 $km*s^{-1}$. SIMBAD reports the galactic coordinates of Cassiopeia A to be $(l,b)=(111.734751^{\circ}, -2.129568^{\circ})$.

The system of two equations described above results in a galactocentric radius $R$ of 11 kpc, which is greater than the galactocentric radius of the LSR. Therefore, there is no Kinematic Distance Ambiguity, and the system of two equations yields only one plausible value for $d$: 4.5 kpc. The other value is -10.6 kpc, which is negative, suggesting that this value is not physically meaningful. Although the sign of the radial velocity of NGC 6309 was changed to be positive, the sign of the radial velocity of Cassiopeia A remained negative, as this was the sign which produced physically meaningful results.

\begin{figure}[ht]
\centering
\includegraphics[width=1.0\textwidth]{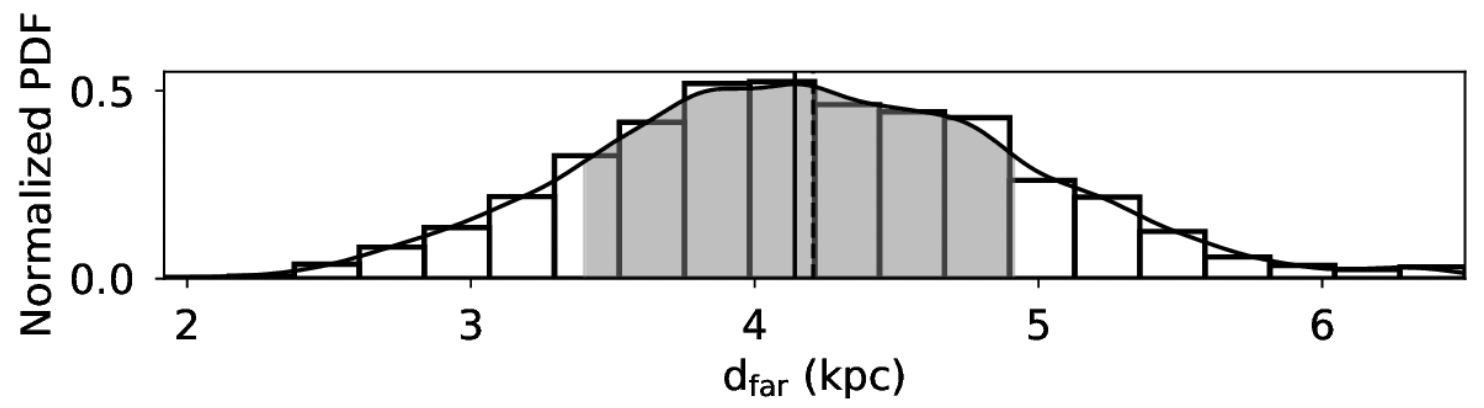}
\caption{\label{fig:Cassiopeia_A_distance.png} The Probability Distribution Function (PDF) generated by the Wenger 2018 software estimates the distance to Cassiopeia A to be between 3.40 kpc and 4.93 kpc with 68.3\% confidence.}
\end{figure}

The Monte Carlo software estimates a distance of 4.14 +0.79/-0.74 pc, where the uncertainty indicates the 68.3\% confidence interval. The Probability Distribution Function (PDF) for the distance to Cassiopeia A is shown in figure \ref{fig:Cassiopeia_A_distance.png}.

The value produced by the Wenger 2018 software differs from the value produced by the system of two equations by 8\%. This difference is within the Monte Carlo simulations uncertainty interval, suggesting that the two methods produce values which are mostly, but not perfectly, in agreement with each other.

The distance value found through the Zhu 2013 method and the Wenger 2018 method differ from Asgekar 2013's distance value by 33\% and 22\%, respectively. The distance reported by Asgekar 2013 falls within the 68.3\% confidence interval of the PDF. This suggests that the KDMs are imperfect but moderately accurate. 

One possible source of error in this distance estimate is that Cassiopeia A has slow but non-negligible proper motion, as described by Kamper and van den Bergh 1976. Additionally, Cassiopeia A is a supernova remnant, not a planetary nebula, and therefore the distribution of proper motions used by the Wenger 2018 software may be slightly inaccurate for supernova remnants.

\subsection{Distance-Dependent Features of NGC 6309}

The distance to NGC 6309 allowed the authors of this paper to determine a variety of intrinsic properties of the object, such as its diameter, age, absolute magnitude, and luminosity.

Rubio et al 2014 and Vázquez et al 2014 report the angular radius $\theta$ of NGC 6309 to be approximately 32 arcseconds. We use the trigonometric equation

\begin{equation}
\label{eq:radius}
   sin(\theta) = \frac{r}{d} ,
\end{equation}

where $r$ is the radius of the object in parsecs and $d$ is the heliocentric distance to the object in parsecs. We calculate the radius to be 0.14 pc, and therefore the diameter of NGC 6309 is 0.28 pc.

Newton's first law of motion suggests that, after being emitted from its central star, NGC 6309 has been expanding at a constant velocity. Therefore, we approximate its expansion as constant for the entirety of its lifespan. Rubio et al 2014 report an expansion velocity of 5 $km*s^{-1}$ for NGC 6309. Using the radius and expansion velocity, we find a kinematic age of 27,000 years for NGC 6309.

The New General Catalogue reports an apparent magnitude of 11 for NGC 6309.\footnote{The New General Catalogue was accessed via webpage created by the Students for the Exploration and Development of Space (SEDS). The hyperlink for the webpage is: http://spider.seds.org/ngc/ngc.cgi?CatalogNumber=NGC+6309} The distance modulus equation states that:

\begin{equation}
\label{eq:modulus}
   m - M = 5log(d)-5,
\end{equation}

where $m$ is the apparent magnitude, $M$ is the absolute magnitude, and $d$ is the distance to the object. We calculate $m=-1.6$ for NGC 6309. Ciardullo et al 2002 found that the brightest hypothetical planetary nebulae can possess an absolute magnitude of $M = -4.47 \pm 0.05$; a value of -1.6 is well within this limit.

The following equation connects the luminosity of an object in Watts with the bolometric magnitude (i.e. absolute magnitude). The equation is:

\begin{equation}
\label{eq:watts}
   M_{bol} = -2.5*log(\frac{L}{3.0128*10^{28}W})
\end{equation}

We calculate the luminosity of NGC 6309 to be $1.3*10^{29}$ W, which is approximately three orders of magnitude brighter than our sun.

We present these calculations to demonstrate the relative ease at which intrinsic features of a planetary nebula can be calculated when the distance to the object is known.

\section{Conclusion}

We calculate the distance to NGC 6309 (Box Nebula) to be 4.1 +0.29 / -0.38 kpc. The distance allows us to determine other important properties of the nebula, such as its radius, age, absolute magnitude, and luminosity.

One possible source of uncertainty in our calculations, in addition to the sources already discussed, is interstellar reddening. The absorption and subsequent reemission of light by the interstellar medium may create an overly redshifted spectrum of NGC 6309. Therefore, the radial velocity estimates may possibly be too large if the spectra was subject to interstellar extinction. We encourage future papers to investigate this potential source of error.

We also hope that future papers will utilize this revised distance estimate in order to calculate other intrinsic features of NCG 6309, such as its mass, density, and temperature.

Additionally, we encourage future astrophysicists to perform this method on other planetary nebulae in the New General Catalogue. Many NGC objects only have estimated distances recorded in databases such as SIMBAD.

If astronomers use the KDM to find distances to hundreds of NGC objects, the astronomical community could use the aggregate data to calculate updated estimates of formation rates of PNe and galactic distribution of PNe. These future discoveries may help astrophysicists acquire a deeper understanding of stellar life cycles in the Milk Way galaxy.

\section{Appendix: The Local Standard of Rest}
Our solar system does not perfectly follow the galactic rotation curves describes in equations \ref{eq:circular velocity} or \ref{eq:Reid}. Instead, the solar system has proper motion relative to what the equations would predict. However, the proper motion occurs at a known velocity, both in direction and in magnitude, whose values are reported in Wenger 2018.

The Local Standard of Rest (LSR) is the point in space which the solar system would follow if it had no proper motion and perfectly adhered to the expected galactic rotation curve. In particular, the LSR is at the solar system's galactocentric radius from the center of the galaxy and rotates at a rate of 220 $km*s^{-1}$ in the direction of $(l,b)=(90^{\circ},0^{\circ})$

When astronomers observe an object's redshift, a portion of that redshift is due to the object moving away from the solar system relative to the LSR, but another portion of the redshift is due to the solar system approaching the object relative to the LSR. Only the first of these values is relevant to the Kinematic Distance Method. Therefore, the radial velocities of objects reported in this paper are the radial velocities of objects relative to the LSR.

\section{Acknowledgements}

We would like to thank Melanie Crowson for her helpful feedback and suggestions while writing this paper.

Thank you to Jack Layton for helping us utilize and understand the Monte Carlo software.

Thank you to Timothy DeLisle for his useful feedback and suggestions about what problems to be aware of.

Thank you to Thurburn Barker for his help conceptualizing the physics behind this project.

Thank you to Amanda Peake for her help in understanding the mathematics behind this paper.

\end{document}